\documentclass[floats,twocolumn,prb]{revtex4}
\usepackage{graphicx}
\usepackage{dcolumn}
\usepackage{bm}
\usepackage{mathrsfs}

\begin{document}
\preprint{APS/123-QED}

\title{Doublon-holon excitations split by Hund's rule coupling within
the orbital-selective Mott phase}

\author{Yuekun Niu$^1$}
\author{Jian Sun$^2$}
\author{Yu Ni$^1$}
\author{Jingyi Liu$^1$}
\author{Yun Song$^1$}\emph{}
\thanks{yunsong@bnu.edu.cn}
\author{Shiping Feng$^1$}

\affiliation{%
$^1$Department of Physics, Beijing Normal University, Beijing 100875, China}%

\affiliation{%
$^2$Beijing National Laboratory for Condensed Matter Physics,Institute of Physics, Chinese Academy of Sciences, Beijing 100190, China}%

\date{\today}

\begin{abstract}
Multiorbital interactions have the capacity to produce
an interesting kind of doublon-holon bound state that consists of a
single-hole state in one band and a doubly-occupied state
in another band. Interband doublon-holon pair excitations in the
two-orbital Hubbard model are studied by using dynamical
mean-field theory with the Lanczos method as the impurity solver.
We find that the interband bound states may provide several
in-gap quasiparticle peaks in the density of states of the narrow
band in the orbital-selective Mott phase with a small Hund's
rule coupling ($J$). There exists a corresponding energy relation
between the in-gap states of the narrow band and the peaks in the
excitation spectrum of the doublon for the wide band.
We also find that the spin flip and pair-hopping Hund interactions
can divide one quasiparticle peak into two peaks, where
the splitting energy increases linearly with increasing $J$.
Strong Hund's rule coupling can move the interband doublon-holon pair
excitations outside the Mott gap and restrict the bound
states by suppressing the orbital selectivity of the doubly-occupied and
single-hole states.
\end{abstract}


\maketitle


\section{INTRODUCTION}
\label{sec:INTR}

The cooperative effect of electron-electron interactions
and orbital degeneracy gives rise to a variety of intriguing
phenomena in strongly correlated multiorbital systems.
\cite{Imada-1998,Kotliar-2006,Rohringer-2018}
The interactions in a multiorbital Hubbard model
typically consist of three components:
an intraorbital Hubbard interaction $U$, an interorbital
Coulomb repulsion $U'$, and the Hund's rule coupling $J$.
Theoretical studies demonstrate that the effective Coulomb
repulsion is increased by a finite Hund's rule coupling
$J$, which results in a strong reduction in the critical
correlation $U_c$ of the Mott transition.
\cite{Werner-2007,Nevidomskyy-2009,Georges-2013}
Owing to the effect of the Hund's rule coupling, which may
greatly suppress interorbital charge fluctuations,
an orbital-selective Mott transition (OSMT) will occur,
where the carries on a subset of orbitals become localized
while the others remain metallic. \cite{Anisimov-2002}

Four factors may lead to an OSMT in multiorbital systems:
(1) The	bandwidth difference plays an essential role in the occurrence of
the OSMT, which	has	been verified by some dynamical
mean-field theory (DMFT) investigations;
\cite{Koga-2004,Koga-2005,deMedici-2005,Song-2005}
(2) The crystal field splitting reduces the orbital degeneracy
to induce the OSMT;
\cite{Werner-2007,deMedici-2009,Song-2009,Jakobi-2013}
(3) The next nearest-neighbor (NN) hopping breaks the particle-hole
symmetry at half filling that also benefits the emergence of
the OSMT; \cite{YKNiu-2018}
(4) The Hund's rule coupling $J$ promotes the OSMT at
half-filling by strongly suppressing the coherence scale
to block the orbital fluctuations.
\cite{Georges-2013,JSun-2015,deMedici-2011,Liebsch-2005}

The two-orbital Hubbard model is the minimal
theoretical model used to study the OSMT.
\cite{Koga-2004,Koga-2005,deMedici-2005,Song-2005,
Costi-2007,deMedici-2009,Song-2009,YKNiu-2018,JSun-2015,
deMedici-2011,Liebsch-2005,Medici-2011,Jakobi-2013}
In the vicinity of the OSMT, a finite $J$ can lead to
fundamentally different low-energy behavior in the
two-orbital Hubbard model. \cite{Greger-2013}
A very recent DMFT study \cite{Fernandez-2018} found
an interesting kind of doublon-holon bound state in the
two-orbital Hubbard model when the OSMT occurs.
Because the quasiparticle peak of the doublon-holon pair
excitation is locked at the Fermi energy when $U=U'$, the
OSMT cannot occur, regardless of the difference in the bandwidths
of the two orbitals.
\cite{Fernandez-2018}

A doublon (holon) is an excitation in which one particle is added
to (removed from) a lattice site with average integer filling.
The unique properties of a Mott insulator
require the doublon and holon to form a
bound state.
%
%
\cite{XJHan-2016,Phillips-2010}
For the single-band Hubbard model, sharp subpeaks have been found at the
inner edges of the Hubbard bands in the metallic phase close to the
Mott transition.
\cite{Leigh-2009,Yamaji-2011,Zhou-2014,Lee-2017,Lee-2-2017}
However, the existence of subpeaks in
the insulating phase is still a matter of debate.
\cite{Nishimoto-2004,Gull-2010,Granath-2014,Lee-2017,Lee-2-2017}

In a multiorbital system, there exists a specific relationship between
the doublon-holon bound state and the OSMT.
The orbital-selective Mott phase (OSMP) between the metallic and 
insulating phases provides
a new perspective for investigating the properties of
doublon-holon pair excitations.
Multiorbital interactions may also have the capacity to
introduce different types of doublon-holon pairs.
Very recently, an interesting kind of doublon-holon bound state
was found in the OSMP of the two-orbital Hubbard model without
the interaction terms for the Hund's rule coupling $J$.
\cite{Fernandez-2018}
This doublon-holon pair excitation consists of a single-hole state
in one band and a doubly-occupied state in the other band,
which is called an interband doublon-holon bound state.
The interband doublon-holon pair excitations provide
quasiparticle peaks in the narrow band (NB) only in the
presence of a coherent metallic resonance in the wide band (WB).
\cite{Fernandez-2018}
However, the above findings are mainly based on the assumption that
$J=0$. Hence it is still unclear how the Hund's rule spin
exchange influences the formation of the interband doublon-holon
pair excitations.

In this paper, we study the effect of Hund's rule coupling on the
doublon-holon bound states in the two-orbital
Hubbard model by using DMFT with the Lanczos method as the
impurity solver.
We find that some in-gap quasiparticle peaks can appear in
the density of states (DOS) of the insulating NB for the
OSMP with a smaller Hund's coupling and bandwidth ratio.
These spectral features indicate the occurrence of the interband
doublon-holon bound states, and the orbital selectivity of the
doubly-occupied state and single-hole state can be found by
investigating the excitation spectra of the doublon and holon.
In an OSMP, Hund's coupling can split one low-energy quasiparticle
peak into two subpeaks, and the energy gap between the two
subpeaks is $2J$.  The splitting of the
quasiparticle peak is mainly caused by the spin flip and pair-hopping
Hund interactions.

Suppression effects on the excitation spectra of the doublon and
holon are found when we increase the Hund's coupling.
In addition,
the distance from the Fermi level to the nearest peak increases
linearly with increasing $J$. As a result, the quasiparticle peaks of
the interband doublon-holon pairs may be moved outside the Mott gap,
and hence are not easily identified from the high-energy excitations of the
Hubbard bands.
We also find that the in-gap spectral features disappear
completely in the fully insulating phase.

This paper is organized as follows.
In Sec. II, we introduce
the theoretical model and the DMFT numerical approach.
In Sec. III, we calculate the spectral function and optical conductivity
to show the influence of Hund's rule coupling on the doublon-holon
bound states.
We discuss the conditions for the occurrence of in-gap quasiparticle
excitations and the interband feature of the doublon-holon pair excitations.
The principal findings of this paper are summarized in Sec. IV.

\section{Two-orbital Hubbard model and dynamical mean-field method}

We consider the Hamiltonian of the two-orbital Hubbard
model,

\begin{eqnarray}
    H&=&-\sum_{\langle ij \rangle l \sigma}t_{l}d^{\dag}_{il\sigma}d_{jl\sigma}
    -\mu\sum_{il\sigma}d^{\dag}_{il\sigma}d_{il\sigma}
    \nonumber\\
    &&+\frac{U}{2}\sum_{il\sigma}n_{il\sigma}n_{il\bar{\sigma}}+\sum_{i\sigma\sigma'}
    (U'-\delta_{\sigma\sigma'}J)n_{i1\sigma}n_{i2\sigma'}
    \nonumber\\
    &&+\frac{J}{2}\sum_{i,l\neq l',\sigma} d^{\dag}_{il\sigma}d^{\dag}_{il\bar{\sigma}}d_{il'\bar{\sigma}}d_{il'\sigma}
    \nonumber\\
    &&+\frac{J}{2}\sum_{i,l\neq l',\sigma\sigma'} d^{\dag}_{il\sigma}d^{\dag}_{il'\sigma'}d_{il\sigma'}d_{il'\sigma},
\label{Eq:TOHub}
\end{eqnarray}
where $\langle ij \rangle$ represents the NN
sites on a Bethe lattice, $d^{\dag}_{il\sigma}$ ($d_{il\sigma}$)
is the electron creation (annihilation) operator for the orbital
$l$ (=1 or 2) at site $i$ with spin $\sigma$, and
$n_{il\sigma}=d^{\dag}_{il\sigma}d_{il\sigma}$ represents the
electron occupation of the orbital $l$ at site $i$. $
t_l$ denotes the NN intraorbital hopping in
orbital $l$, $U$ ($U'$) corresponds to the intraorbital
(interorbital) interactions, and $J$ is the Hund's rule coupling.
The last two terms represent the pair-hopping and spin flip
Hund interactions, respectively. For systems with spin rotation
symmetry, the relationship $U=U'+2J$ should be kept.

Considering the semicircular DOS of the Bethe
lattice, the onsite component of the Green's function of
each orbital ($G_{ii}^{(l)}(i\omega_n)=\sum_k G_l(k, i\omega_n)$)
satisfies a simple self-consistent relation,
\begin{equation}
\{g^{(l)}_{0}(i\omega_{n})\}^{-1}
=i\omega_{n}+\mu-t^{2}_{l}G_{ii}^{(l)}(i\omega_{n}),
\label{g0-G}
\end{equation}
where $g_{0}$ is the noninteracting Green's function.
 \cite{Georges-1996}

In a DMFT procedure, the lattice Hamiltonian (Eq.~(\ref{Eq:TOHub}))
needs to be mapped onto an impurity model with fewer degrees of
freedom,
\begin{eqnarray}
    H_{imp}&=&\sum_{ml\sigma}\epsilon_{ml\sigma} c^{\dag}_{ml\sigma}c_{ml\sigma}
    -\mu\sum_{l\sigma}d^{\dag}_{l\sigma}d_{l\sigma}
    \nonumber\\
    &&+\sum_{ml\sigma}V_{ml\sigma}(c^{\dag}_{ml\sigma}d_{l\sigma}+d^{\dag}_{l\sigma}c_{ml\sigma})
    \nonumber\\
    &&+\frac{U}{2}\sum_{l\sigma}n_{l\sigma}n_{l\bar{\sigma}}+\sum_{\sigma\sigma'}
    (U'-\delta_{\sigma\sigma'}J)n_{1\sigma}n_{2\sigma'}
    \nonumber\\
    &&+\frac{J}{2}\sum_{l\neq l',\sigma}d^{\dag}_{l\sigma}d^{\dag}_{l\bar{\sigma}}d_{l'\bar{\sigma}}d_{l'\sigma}\\
    &&+\frac{J}{2}
    \sum_{l\neq l',\sigma}d^{\dag}_{l\sigma}d^{\dag}_{l'\sigma'}d_{l\sigma'}d_{l'\sigma},\nonumber\label{IMP}
\end{eqnarray}
where $\epsilon_{ml\sigma}$ denotes the effective parameter of
the $m$-th environmental bath of orbital $l$,
and $V_{ml\sigma}$ represents the coupling
between the impurity site and its environment baths. The parameters
$\epsilon_{ml\sigma}$ and $V_{ml\sigma}$ are determined by performing
self-consistent DMFT calculations using an impurity solver.

In our study, we employ the Lanczos solver.\cite{Dagotto-1994}
The Green's function $G^{(l)}_{imp}(i\omega_n)$ of the impurity model
can be expressed as,  \cite{Caffarel-1994,Georges-1996,Capone-2007}
\begin{eqnarray}
    G_{imp}^{(l)}(i\omega_n)&=&G_l^{(+)}(i\omega_n)+G_l^{(-)}(i\omega_n),
\end{eqnarray}
where
\begin{eqnarray}
    G_l^{(+)}(i\omega_n)&=&\frac{\langle \phi_0|d_l d_l^{\dag}|\phi_0 \rangle}{i\omega_n-a^{(+)}_0
    -\frac{b^{(+)2}_1}{i\omega_n-a^{(+)}_1-\frac{b^{(+)2}_2}{i\omega_n-a^{(+)}_2-...}}},\\
    G_l^{(-)}(i\omega_n)&=&\frac{\langle \phi_0|d_l^{\dag}d_l|\phi_0 \rangle}
    {i\omega_n+a^{(-)}_0-\frac{b^{(-)2}_1}{i\omega_n+a^{(-)}_1-\frac{b^{(-)2}_2}{i\omega_n+a^{(-)}_2-...}}}.
\end{eqnarray}

\begin{figure}[htbp]
\includegraphics[scale=0.5]{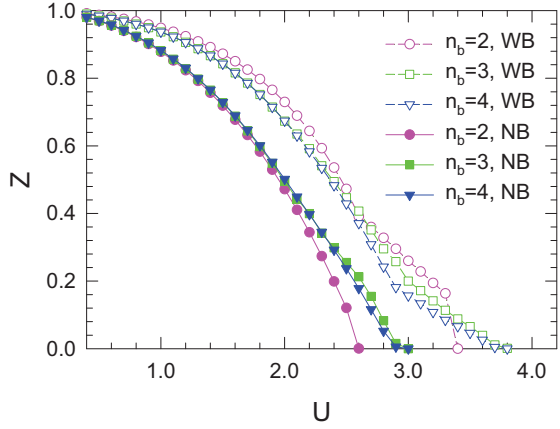}
\caption{(Color online) Effect of the bath size $n_b$ on the critical interactions
$U_c$ the OSMT in the two-orbital Hubbard model.
The interaction dependencies of the quasiparticle weight ($Z$) of the WB
(dashed lines and hollow symbols) and NB (solid lines
and filled symbols) are shown for various
bath sizes: $n_b=2$ (circles), $n_b=3$ (squares), and $n_b=4$ (triangles).
The same critical values of the OSMT, $U_{c1}=3.8$ and $U_{c2}=3.0$, are obtained
for the two cases with different bath sizes $n_b=3$ and $n_b=4$.
The other model parameters are: $t_2/t_1=0.6$, $J=U/4$, and $U=U'+2J$.
The energies are in units of $t_1$.}
\label{fig:ZBth}
\end{figure}

In Eq.~(\ref{Eq:TOHub}), the two orbitals are nonhybridized.
Thus, the self-energy, effective medium functions,
and Green's functions are all diagonal with respect to
the orbitals. Within multiorbital DMFT
calculations, \cite{Arita-2005} the frequency energy is
defined as $\omega_n=(2n+1)\pi/\beta$.
In our calculations, we choose $\beta=512$ to
assure the accuracy of the self-consistency for the Green's
functions,
$G_l(i\omega_{n})=G^{(l)}_{ii}(i\omega_{n})=G_{imp}^{(l)}(i\omega_{n})$,
especially in the low-energy region.
The quasiparticle weights $Z_l$ of different bands can be obtained by,
\begin{equation}
Z_l=(1-\frac{\partial}{\partial\omega}\rm{Re}\it{\Sigma_l(\omega)|_{\omega=0}})^{-1}
\approx (1-\frac{\rm{Im}\it{\Sigma_{l}(i\omega_{0})}}{\omega_{0}})^{-1}.
\end{equation}
Analytic continuation is performed to obtain
the real frequency Green's function $G_l(\omega)$.
\cite{Georges-1996}
We calculate the orbital-resolved DOS by
$\rho_l(\omega)=-\frac{1}{\pi}\rm{Im}\it{G_l(\omega+i\delta)}$,
where $\delta$ is a factor for energy broadening.
The orbital-dependent optical conductivity is
expressed as
\begin{eqnarray}
\label{optical-conductivity-d}
\sigma_l(\omega)&=&\pi\int_{-\infty}^{\infty}d\epsilon D_l(\epsilon)\int_{-\infty}^{\infty}\frac{d\omega^{\prime}}{2\pi}
\rho_l^{(\epsilon)}(\omega^{\prime})\rho_l^{(\epsilon)}(\omega^{\prime}+\omega)
\nonumber\\
&&\times\frac{n_{f}^{(l)}(\omega^{\prime})-n_{f}^{(l)}(\omega^{\prime}+\omega)}{\omega},
\end{eqnarray}
where $n_{f}(\omega)$ is the Fermi function, and
$D_l(\epsilon)=\frac{1}{2\pi t_l}\sqrt{4t_l^2-\epsilon^2}$ is
the semicircular DOS of the Bethe lattice.

\section{Results}

\subsection{Phase diagram of the OSMT}
\label{Sec:PDs}

\begin{figure}[htbp]
\includegraphics[scale=0.42]{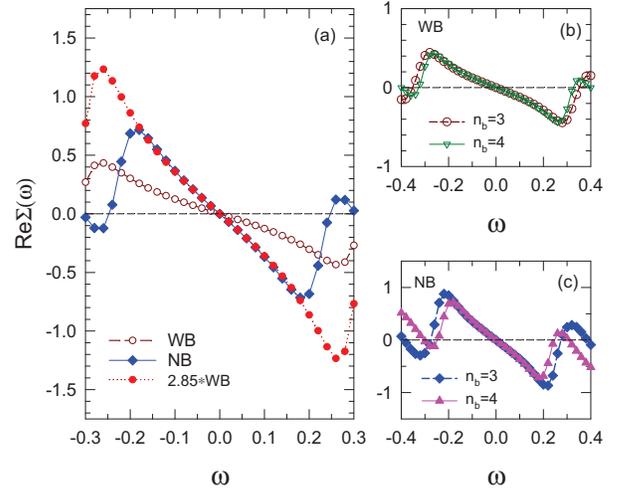}
\caption{
Common low-energy scale induced by Hund's rule coupling in the vicinity of
 the OSMT.
(a) The self-energies of the two bands has the relation of
$\rm{Re}\Sigma_2(\omega)=2.85\rm{Re}\Sigma_1(\omega)$
within the energy region [-0.2, 0.2]
when the two-orbital Hubbard model is close to the OSMT with
$J=0.5$, $U=2$, $U=U'+2J$, and $t_2/t_1=0.5$.
The self-energies of the WB (b) and NB (c) are
almost the same for $n_b=3$ and $n_b=4$
within the corresponding low-energy region.
}
\label{fig:RSE}
\end{figure}

The existence of the OSMT in a nondegenerate
two-orbital Hubbard model is demonstrated by the evolution
of the quasiparticle weight $Z_l$
with increasing interactions $U$ when $t_2/t_1=0.6$ and $J=U/4$,
as shown in Fig.~\ref{fig:ZBth}.
The interaction dependence of $Z_l$ for the cases
with $n_b$=3 and $n_b=4$ are very similar, which gives the same
critical values of the OSMT: $U_{c1}=3.8$ for the WB
and $U_{c2}=3.0$ for the NB.
For the two-orbital Hubbard model with parameters close
to the OSMT, a common low-energy scale is found
when the Hund's rule coupling is strong.
\cite{Greger-2013}
As shown in Fig.~\ref{fig:RSE}(a), the self-energy of the
NB is approximately equal to the product of the self-energy
of the WB and a certain constant, i.e.,
$\rm{Re}\Sigma_2(\omega)\approx \alpha\rm{Re}\Sigma_1(\omega)$,
in the low-energy region [-0.2, 0.2].
The constant $\alpha$ is found to be
2.85 when $U=2$ and $J=0.5$.

\begin{figure}[htbp]
\includegraphics[scale=0.39]{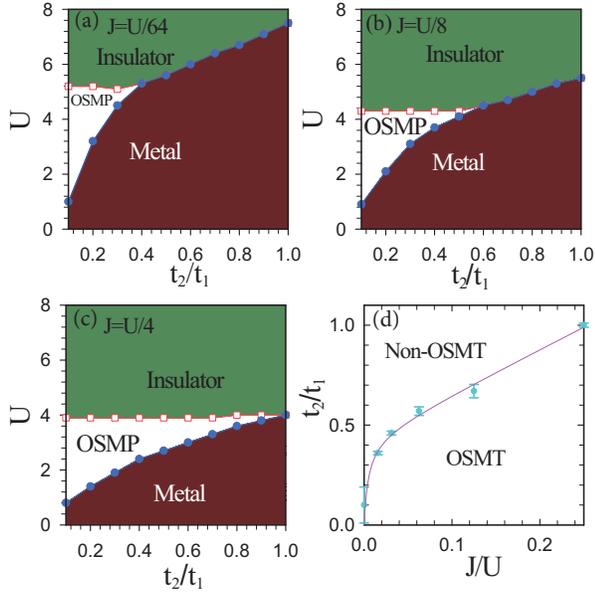}
\caption{Phase diagrams of the two-orbital Hubbard model with various Hund's rule
couplings: $J=U/64$ (a), $J=U/8$ (b), and $J=U/4$ (c).
Both the critical values $U_{c1}$ and $U_{c2}$ for the WB and NB
decrease as the Hund's coupling increases due to the enhancement
in the Coulomb interactions caused by $J$.
When the Hund's coupling is sufficiently strong ($J\geq U/4$), the OSMT can
occur for any bandwidth ratio $t_2/t_1$.
(d) Dependence of the boundary between the OSMT region and non-OSMT region on
the Hund's coupling.
As $J$ decreases, a significant decline in the threshold of the ratio
$t_2/t_1$ is observed, which drops to zero when $J=0$.
Thus, there is no OSMT for any nonzero bandwidth in both
bands when $U=U'$ and $J=0$.}
\label{fig:PhsD}
\end{figure}

In agreement with the prediction of some previous DMFT calculations,
\cite{Liebsch-2005} our study shows that one can accurately
determine the critical points of Mott transitions in the two-orbital
Hubbard model by using the Lanczos solver with a limited bath size.
Therefore, we could comprehensively investigate the influence of different
model parameters on the phase diagram of
the two-orbital Hubbard model, especially the Hund's coupling $J$.

%
%
%
In Fig.~\ref{fig:PhsD}, we compare the phase diagrams of
the two-orbital Hubbard model with different Hund's couplings.
For the cases with a small $J$, the OSMT occurs only if the
orbital difference meets a certain requirement.
For example, the appearance of the OSMP requires $t_2/t_1\leq 0.6$
when $J=U/8$, as shown in Fig.~\ref{fig:PhsD}(b).
However, the OSMP can exist for
any bandwidth ratio when the Hund's rule coupling is
sufficiently strong.
As illustrated in Fig.~\ref{fig:PhsD}(d), a boundary for the
existence of the OSMT is presented, which clearly shows that
the Hund's rule coupling significantly promotes the OSMT.
There is no OSMT for any nonzero bandwidth in both bands when
$J=0$.
%
%
When $J=0$, the quasiparticle peaks of the interband
doublon-holon pairs will be locked at the Fermi level,
leading to the simultaneous appearance of the Mott
transition for both bands, regardless of the difference
in bandwidths. \cite{Fernandez-2018}

\begin{figure}[htbp]
\includegraphics[scale=0.55]{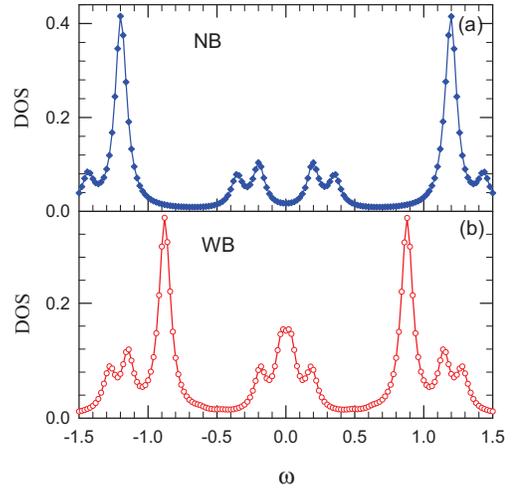}
\caption{(Color online)
DOS showing the low-energy quasiparticle states in the
OSMP of the nondegenerate
two-orbital Hubbard model with a small Hund's rule coupling $J$.
In the low-energy region around the Fermi level, the DOS of the
NB (a) and WB (b) are presented for the OSMP with
$U=5.1$, $J=U/64\approx0.08$, $t_2/t_1=0.2$, and $U=U'+2J$.
The corresponding OSMT critical interactions are $U_{c1}=5.3$ and $U_{c2}=3.2$.
The OSMP is very close to the insulating transition point.
Four quasiparticle peaks are found close to the Fermi level in the NB,
and the corresponding excitations carry energies of
$E=\pm0.17$ and $\pm0.34$, respectively.
The energies are in units of $t_1$, and the energy broadening is $\delta=0.05$.}
\label{fig:DOSJ64}
\end{figure}

\subsection{Quasiparticle excitations in the OSMP}
\label{Sec:Peaks}
%
We find in the OSMP that some low-energy quasiparticle peaks
appear inside the Mott gap of the NB when the
Hund's coupling is sufficiently small.
Fig.~\ref{fig:DOSJ64} shows the DOS of the NB and WB of
the two-orbital Hubbard model in an OSMP.
Four peaks of quasiparticle
excitations are found close to the Fermi level in the NB,
as shown in Fig.~\ref{fig:DOSJ64}(a).
Here, the model parameters are $U=5.1$, $J=0.08$, $t_2/t_1=0.2$, and
$U=U'+2J$. The corresponding Mott critical values for the WB
and NB are obtained as $U_{c1}=5.3$ and $U_{c2}=3.2$,
respectively.
Therefore, the system with $U=5.1$ is in an OSMP,
which is close to the insulating transition point $U_{c1}$.

In the NB,
the four peaks are symmetrically located around the Fermi level,
carrying energies of $E=\pm0.19$ and $\pm0.36$.
It is important to note that the energy splitting between the
two nearby quasiparticle peaks with positive (negative) energy
is $\Delta=0.17$, which is approximately equal to $2J$ ($J$=0.08).
%
%
In the WB, we can also find two low-energy peaks
at the two sides of the center coherent peak, as shown in
Fig.~\ref{fig:DOSJ64}(b). As with the two inner
quasiparticle peaks in the NB, the two low-energy
peaks in the WB carry energies of $E=-0.19$ and $E=0.19$.
This energy association implies that the
quasiparticle bound states may be not orbitally independent.

\begin{figure}[htbp]
\includegraphics[scale=0.49]{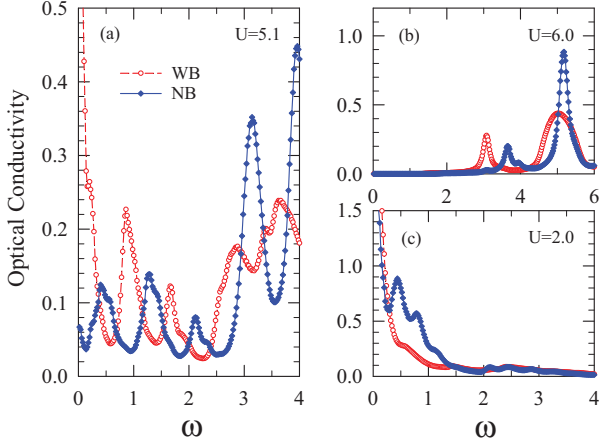}
\caption{(Color online)
Manifestation of the emergence of low-energy bound quasiparticle states
by orbital-resolved optical conductivity.
We compare the optical conductivities of the NB and WB for different
phases: the OSMP (a),
insulating phase (b), and metallic phase (c).
In the OSMP, the low-energy peaks in the NB optical conductivity
(solid line and filled symbols)
indicate the transfer of spectral weight between the quasiparticle peaks
of the low-energy bound excitations appearing in the NB.
These low-energy in-gap excitations completely disappear in the insulating phase.
The other model parameters are the same as those in Fig.~\ref{fig:DOSJ64}.
}
\label{fig:OC}
\end{figure}

Numerous numerical calculations have been carried out for the
two-orbital Hubbard model with various model parameters, and we find that
the energies carried by the quasiparticle peaks do not change with
changes in the bandwidth ratio $t_2/t_1$ when the
Hund's coupling $J$ is fixed.

Low-energy quasiparticle excitations may be observed by
the orbital-resolved optical conductivity
of multiorbital correlated compounds.
In Fig.~\ref{fig:OC}, we present the optical conductivities
of the WB and NB obtained for different phases of
the two-orbital Hubbard model.
The quasiparticle states contribute significantly
to the optical conductivity in the OSMP.
As expected, the optical conductivity of the NB exhibits a significant
feature in the low-energy region, presenting the transfer of
spectral weight between the quasiparticle excitations appearing in the
NB. Meanwhile,
the Drude weight in the optical conductivity of the WB indicates that
the WB is metallic.
In contrast, Drude peaks are shown for both optical conductivities of
the two bands in the metallic phase
(Fig.~\ref{fig:OC}(c)).

The Mott transition occurs in the WB when $U>U_{c1}$,
accompanied by the vanishing of the coherent metallic
resonance and the quasiparticle peaks.
The orbital-dependent optical conductivity can also illustrate
the disappearance of the quasiparticle peaks in the insulating phase,
as shown in Fig.~\ref{fig:OC}(b).
Owing to the absence of low-energy excitations, Mott gaps are
clearly shown in the optical conductivities for both bands
(Fig.~\ref{fig:OC}(b)).
This finding is of great significance for solving the dispute
regarding whether there are
subpeaks in the insulating phase of the single-band
Hubbard model.

\begin{figure}[htbp]
\includegraphics[scale=0.4]{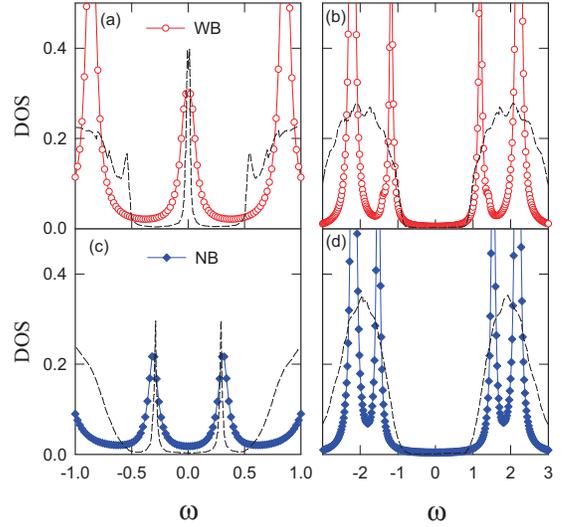}
\caption{(Color online)
Low-energy quasiparticle peaks with broken spin-symmetry.
Left panel: The low-energy DOS of the WB (a)
and NB (c) in the OSMP
when $J=0$, $U=3$, and $U-U'=0.3$.
Two quasiparticle peaks are found in the NB,
located approximately at $\omega=-0.3$ and $\omega=0.3$.
Right panel: The DOS of the WB (b) and NB (d) in the insulating phase
when $J=0$, $U=4.0$, and $U-U'=0.3$.
The quasiparticle peaks in the NB disappear with the vanishing of the
central resonance peak in the WB.
Compared with the results obtained by the DMFT+DMRG (black dashed lines),~\cite{Fernandez-2018}  very good agreement is achieved.
The other model parameters are $t_1=0.5$ and $t_2=0.25$.
}
\label{fig:CpJ0}
\end{figure}

\subsection{Effect of $J$ on the quasiparticle excitations}
\label{Sec:J}

N\'{u}\~{n}ez-Fern\'{a}ndez $et$ $al.$ studied the low-energy bound states in a
simplified two-orbital Hubbard model. \cite{Fernandez-2018}
Without the Hund's coupling terms, this model is still able to show
the OSMT, but the spin rotation symmetry is broken when $U\neq U'$.
The authors found that a finite DOS at the Fermi energy in the WB
is correlated with the emergence of well-defined quasiparticle states
at the excited energy $\Delta=U-U'$ in the insulating NB.
\cite{Fernandez-2018}
For a comparison with their results, we also calculate the DOS of the two-orbital
Hubbard model with $J=0$ and $U\neq U'$.
Our results are in good agreement with the results obtained
by using DMFT with the density-matrix renormalization
group method as the impurity solver, \cite{Fernandez-2018}
as shown in Fig.~\ref{fig:CpJ0}.
Our finding indicates that the splitting of quasiparticle excitations
is caused by Hund's rule coupling. In the NB, as shown in Fig.~\ref{fig:CpJ0}(c),
there are only two quasiparticle peaks when
$J$ is absent.
In addition, there is no quasiparticle excitation in both the WB
and NB for the fully insulating phase, as shown in
Figs.~\ref{fig:CpJ0}(b) and \ref{fig:CpJ0}(d).
N\'{u}\~{n}ez-Fern\'{a}ndez $et$ $al.$ predicted that these quasiparticle excitations
are interband holon-doublon bound states. \cite{Fernandez-2018}

\begin{figure}[htbp]
\includegraphics[scale=0.5]{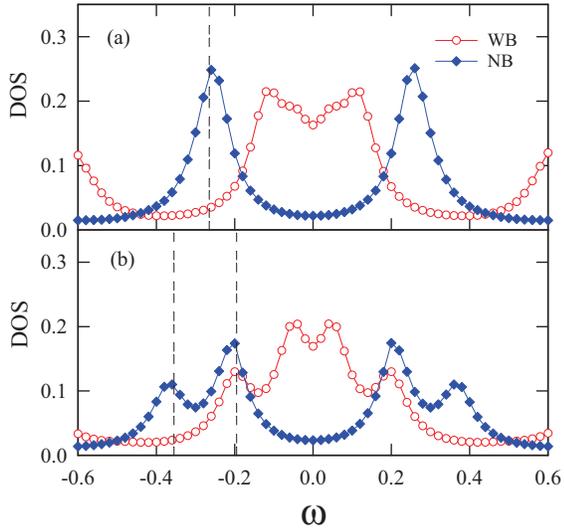}
\caption{(Color online)
Influence of the different interaction terms of the Hund's coupling
Hamiltonian on the quasiparticle excitations:
(a) when only the interorbital density-density Hund's interactions
are considered, and (b) when the spin flip and
pair-hopping Hund interactions are also included.
The positions of the in-gap quasiparticle peaks in the NB
are indicated by the dashed lines in the negative energy
region. It is obvious that the splitting of the quasiparticle
excitations is mainly driven by the spin flip and
pair-hopping Hund interactions.
The model parameters are the same as those in Fig.~\ref{fig:DOSJ64}.
}
\label{fig:Jsfph}
\end{figure}

To understand the effects of the different terms in the
Hund interaction Hamiltonian on the energy splitting of
quasiparticle excitations, we compare the low-energy DOS
of two different models with different Hund interactions,
as shown in Fig.~\ref{fig:Jsfph}.
When only the interorbital density-density Hund interactions
are considered, there are only two quasiparticle peaks
in the NB, which are located at $\omega=\pm 0.265$.
We suppose that the energies of the quasiparticle peaks
may be determined by $D=U-U'+J$.
The interaction parameters are $U-U'$=0.16, $J=0.08$, and
$U=5.1$ ($U-U'=2J$). Our prediction is in good agreement
with the finding of Ref.~[22] ($J=0$ and $D=U-U'=0.3$),
where the two peaks carry energies of $0.3$
and $-0.3$, as shown in Fig.~\ref{fig:CpJ0}(c).

\begin{figure}[htbp]
\includegraphics[scale=0.53]{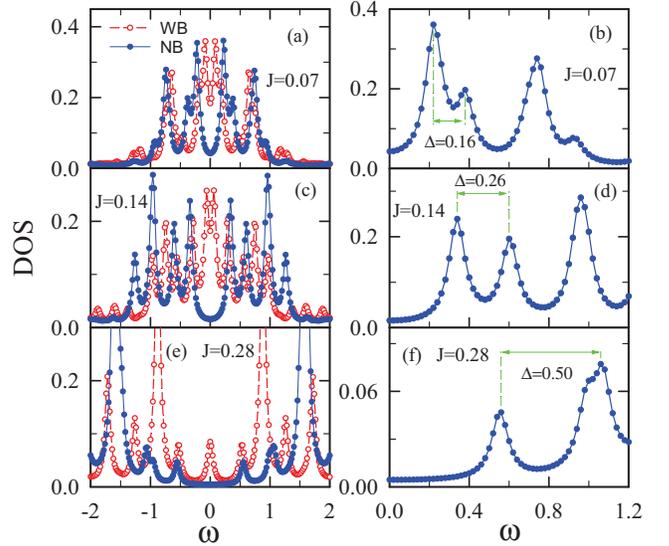}
\caption{(Color online)
Quasiparticle excitations are split by Hund's rule coupling
in the OSMP.
Left panels: The DOS of the WB and the NB
for different Hund's couplings:
$J=0.07$ (a), $J=0.14$ (c), and $J=0.28$ (e).
Right panels: A one-to-one correspondence with the left panels. The energy
splitting between the quasiparticle peaks is shown in the low-energy
DOS of the NB.
The two-orbital Hubbard model remains in the OSMP with $t_2/t_1=0.2$
and $U=4.5$.
The positions of the quasiparticle peaks increase linearly with increasing
Hund's rule coupling.
A linear dependence of the energy splitting between the peaks on $J$
is also found.
}
\label{fig:SptJ}
\end{figure}

Four quasiparticle peaks appear in the NB when the
influences of the full Hund interactions are considered.
As shown in Fig.~\ref{fig:Jsfph}(b), the energies carried
by the four quasiparticles in the NB are
0.36, 0.19, -0.19, and -0.36.
Our results indicate that the energy splitting of the
interband holon-doublon bound states is mainly caused by the
spin flip and pair-hopping Hund interactions.
As a special type of double-hopping term, the pair-hopping
Hund's coupling can move two electrons from one orbital to
another simultaneously, which contributes to the occurrence
of interband doublon-holon pairs and the transition
between different interband bound states.
Similarly, the spin-flip exchange interaction also has a
significant effect on the interorbital doublon-holon bound
states because it represents a particular kind of
double-hopping term between the two orbitals. The transverse
(spin flip and pair-hopping) Hund's couplings enhance the
electronic interactions and spin fluctuations, resulting in
the splitting of the doublon-holon excitations.
The interplay between the split doublon-holon bound
states and the dependence of the splitting energy on
effective doublon-holon pair interactions require further
investigations.

As mentioned
in the previous subsection, there might also exist some
low-energy quasiparticle peaks in the WB. However,
the WB is in the metallic phase, and there is a
resonance peak at the Fermi level. Therefore, distinguishing
the low-energy quasiparticle excitations with the central
resonance peak is difficult.
We find that the overlap between the interband bound states
with the resonance peak is reduced when the
spin flip and pair-hopping Hund interactions are included.
Also shown in Fig.~~\ref{fig:Jsfph}(b), two quasiparticle
peaks are found at $\pm 0.19$ in the WB, which have the
same energies as the two inner peaks in the NB.

We further study the relation between the energy splitting $\Delta$ and $J$.
We focus on the two-orbital Hubbard model with full Hund's rule
coupling.
In Fig.~\ref{fig:SptJ}, from top to bottom, Hund's
coupling increases from $J=0.07$ to $J=0.14$ and finally to $J=0.28$,
and the intraorbital interactions are fixed as $U=4.5$.
In the left panel, the DOS of the WB and NB clearly show
that the two-orbital Hubbard model remains in the OSMP with the
change in the Hund's rule coupling. Correspondingly, in the right
panel, we indicate the energy splitting $\Delta$ between the two nearby
quasiparticle peaks in the NB for the three cases with different $J$.
It is shown that the splitting energy increases linearly with
increasing $J$, satisfying the relation $\Delta=2J$.

In addition, our study also suggests that there also exists a linear relationship
between the position of the quasiparticle peak and the Hund's rule coupling $J$.
When the Hund's coupling increases from $J=0.14$ to $J=0.28$, the energy carried
by the inner peak is found to increase from $\omega=0.26$ to $\omega=0.50$.

It is worth noticing that a small bandwidth ratio $t_2/t_1$ is
an essential condition for the emergence of the low-energy
quasiparticle peaks in the OSMP of the two-orbital Hubbard model.
Many earlier DMFT studies focused on the OSMT with a bandwidth
ratio of $t_2/t_1=0.5$ only, which may be the main reason why the
feature of the quasiparticle peaks was missing.
Based on the phase diagrams shown in Fig.~\ref{fig:PhsD}, to find the OSMP
for the cases with $t_2/t_1=0.5$, the Hund's rule coupling must be larger
than 0.5. The quasiparticle peaks are predicted to appear in
the high-energy region ($U-U'>1.0$). Thus, it would be difficult to
distinguish them from the excitations in the Hubbard bands.

\subsection{Excitation spectra of the doublon and the holon}
\label{Sec:CF}

The spectrum function has well-defined quasiparticle
peaks in the low-energy region when the two-orbital
Hubbard model is in the OSMP with a small Hund's rule
coupling. To characterize the feature of these
quasiparticle excitations, we focus primarily on the
orbital-resolved excitation
spectrum of the doublon $D_{l}(\omega)$, which is defined as
\begin{equation}
D_{l}^{(-)}(\omega)=-\frac{1}{\pi}{\rm{Im}} \langle b^{\dag}_{l}
(\omega-H+i\delta^{+})^{-1}b_{l} \rangle,
\end{equation}
where the doublon operator \cite{Lee-2017}
$b^{\dag}_{l}=n_{l\downarrow}d^{\dag}_{l\uparrow}$
creates a doubly-occupied state in orbital $l$.

\begin{figure}[htbp]
\includegraphics[scale=0.4]{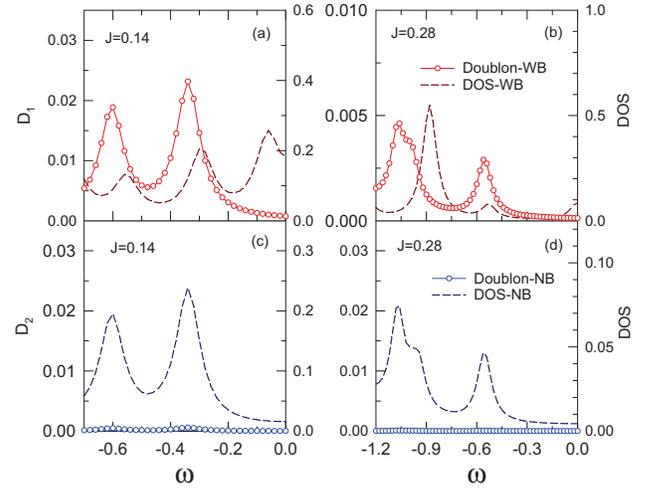}
\caption{(Color online) Orbital selectivity of the doubly-occupied
states in the OSMP.
The excitation spectra of the doublon in the WB ((a) and (c))
and NB ((b) and (d)) are shown by the solid lines with
hollow symbols,  for the OSMP with different Hund's
rule coupling $J=0.14$ and $J=0.28$.
For a comparison, the DOS of the two bands are
also presented by the dashed lines.
The doublon spectrum of the WB is significantly larger than that of the
NB, indicating that the doubly-occupied states prefer to stay
in the WB.
The energies of the in-gap peaks in the DOS of the NB correspond to the
positions of the peaks in the doublon spectrum.
The other model parameters are $U=4.5$, $t_2/t_1=0.2$, and $U=U'+2J$.}
\label{fig:A-DH}
\end{figure}

The excitation spectrum of the doublon within the negative
low-energy region is plotted in Fig.~\ref{fig:A-DH} for
the two-orbital Hubbard model with different Hund's
coupling: $J=0.14$ (left panel) and $J=0.28$ (right panel).
Obviously, the doublon spectrum function of the WB
is much stronger than that of the NB, which
suggests that the doubly-occupied states prefer to stay in
the WB. In addition, the excitation spectrum is found
to decrease with increasing Hund's coupling, which indicates
that strong Hund's coupling suppresses
the orbital selectivity of the doubly-occupied state.

Moreover, a specific correlation between the two bands is
also observed for the first time, where the energies
carried by the quasiparticle excitations in the NB
are determined by the positions of the peaks shown in the
excitation spectrum of the doublon of the WB.
As shown in Figs.~\ref{fig:A-DH}(a) and \ref{fig:A-DH}(c),
in the negative energy region when $U=4.5$ and $J=0.14$,
there are two peaks in the excitation spectrum of the doublon of
the WB (red solid line with circles), which carry
energies of $\omega=-0.32$ and $\omega=-0.59$, respectively.
The DOS of the NB also has two peaks (blue dashed line),
and the positions of these two peaks correspond exactly
to the peaks in the doublon spectrum of the WB.

This investigation provides insight into the intrinsic orbital-selective
characteristics of the doubly-occupied states. Furthermore, the
interorbital correlation between the doublon spectrum and DOS implies
that the quasiparticle excitation should be formed by the doublon and holon
in different orbitals, which is just the interband doublon-holon boud
state. \cite{Fernandez-2018}
We predict that the interband doublon-holon pair with negative energy
should consist of a doublon in the WB and a hole in the NB.
To confirm this hypothesis, we also need to further study the
orbital selectivity of the single-hole state.

Correspondingly, the excitation spectrum of the holon $H_{l}^{(+)}(\omega)$
can be expressed by the following equation:
\begin{equation}
H_{l}^{(+)}(\omega)=-\frac{1}{\pi}{\rm{Im}} \langle h_{l}
(\omega-H+i\delta^{+})^{-1}h^{\dag}_{l} \rangle,
\end{equation}
where $h^{\dag}_{l}=(1-n_{l\downarrow})d_{l\uparrow}$ presents
the holon operator\cite{Lee-2017} of orbital $l$.

\begin{figure}[htbp]
\includegraphics[scale=0.47]{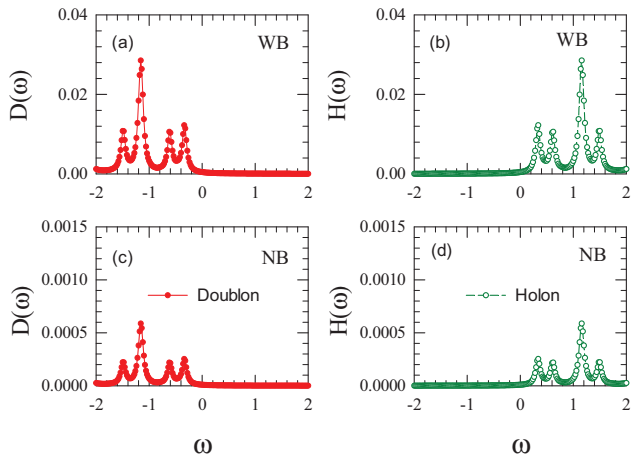}
\caption{(Color online) Excitation spectra of the doublon
(left panels) and holon (right panels) for the WB and the NB
in the OSMP with $J=0.14$, $t_2/t_1=0.2$, $U=4.5$, and $U=U'+2J$.
The spectrum weight of the doublon (holon) of the wide band is approximately 2
orders higher than the corresponding spectrum weight of the NB.
}
\label{fig:LnDH-J}
\end{figure}

In a half-filled system with particle-hole symmetry,
the holon spectrum function $H_{l}^{(+)}(\omega)$ is
not independent.
Based on the particle-hole transformation,
we can find that there is an asymmetric relationship
$D_l^{(-)}(-\omega)=H_{l}^{(+)}(\omega)$ under the transition
$\omega\rightarrow -\omega$.

In Fig.\ref{fig:LnDH-J}, we compare the excitation spectra of
the doublon and the holon in the OSMP.
As expected, the excitation spectrum of a hole in the positive
energy region matches the spectrum of the doubly-occupied state
in the negative energy region.
A basic feature of the interband doublon-holon bound state
is found, where the pair excitation also shows an asymmetric
relation under the energy transition $\omega\rightarrow -\omega$.
The interband doublon-holon pair with positive (negative) energy consists
of a holon (doublon) in the WB and a doublon (holon) in
the NB. These theoretical predictions need to be tested
and verified by experiments. Additionally, the relationship
between the effective doublon-holon pair interaction and Hund's
rule coupling in the multiorbital Hubbard model still needs to be
explored by further research.

\section{Conclusion}
\label{Sec:Con}
 We study the effect of Hund's rule spin
exchange on the doublon-holon pair excitations in the two-orbital
Hubbard model by using DMFT with the
Lanczos method as the impurity solver.
Our calculations show that low-energy quasiparticle peaks occur
in the DOS of the OSMP if
both the Hund's rule coupling $J$ and the bandwidth ratio $t_2/t_1$
are small enough.
These low-energy excitations are the interband doublon-holon
bound states, in which the doublon is located in one band while the
holon is in the other band.

The spin-flip and pair-hopping Hund interactions can divide
one quasiparticle peak into two peaks. The linear relation
$\Delta=2J$ has been confirmed between the energy gap and
the Hund's rule coupling.
In addition, the energies carried by the quasiparticle peaks are
also controlled by the Hund's rule coupling.

There exists a direct correspondence between the energies
of the quasiparticle peaks in one band and the positions of the
peaks in the excitation spectrum of the doublon for the other band.
Our study demonstrates that the interband
doublon-holon pair with positive (negative) energy consists
of a holon (doublon) in the WB and a doublon (holon) in the
NB.

When the Hund's rule coupling is strong, the interband doublon-holon pair
excitations are suppressed with a significant reduction in the
excitation spectra of the doublon and the holon.
In addition, the low-energy quasiparticle peaks are moved to Hubbard
bands and hence are inefficiently identified.
The interband doublon-holon bound states disappear completely in the
fully insulating phase.

\section*{Acknowledgments}
The computational resources utilized in this research were provided
by Shanghai Supercomputer Center. The work is supported by the
the National Natural Science Foundation of China (NSFC),
under Grants  Nos. 11174036, and 11474023.
SF is supported by the National Key Research and Development Program of
China under Grant No. 2016YFA0300304, and NSFC under Grant
Nos. 11574032 and 11734002.

\bibliography{apssamp}

\begin{thebibliography}{9}\label{sec:TeXbooks}%

\bibitem{Imada-1998}
M. Imada, A. Fujimori, and Y. Tokura,
Rev. Mod. Phys. {\bf 70}, 1039 (1998).

\bibitem{Kotliar-2006}
G. Kotliar, S. Y. Savrasov, K. Haule, V. S. Oudovenko,
O. Parcollet, and C. A. Marianetti,
Rev. Mod. Phys. {\bf 78}, 865 (2006).

\bibitem{Rohringer-2018}
G. Rohringer, H. Hafermann, A. Toschi, A.A. Katanin, A.E. Antipov,
M.I. Katsnelson, A.I. Lichtenstein, A.N. Rubtsov, and K. Held,
Rev. Mod. Phys. {\bf 90}, 025003 (2018).

\bibitem{Werner-2007}
P. Werner and A. J. Millis,
Phys. Rev. Lett. {\bf 99}, 126405 (2007).

\bibitem{Nevidomskyy-2009}
A. H. Nevidomskyy and P. Coleman,
Phys. Rev. Lett. {\bf 103}, 147205 (2009).

\bibitem{Georges-2013}
A. Georges, L. de' Medici, and J. Mravlje,
Annu. Rev. Condens. Matter Phys. {\bf 4}, 137 (2013).

\bibitem{Anisimov-2002}
V. I. Anisimov, I. A. Nekrasov, D. E. Kondakov, T. M. Rice, and M. Sigrist,
Eur. Phys. J. B {\bf 25}, 191 (2002).
\bibitem{Koga-2004}
A. Koga, N. Kawakami, T. M. Rice, and M. Sigrist,
Phys. Rev. Lett. {\bf 92}, 216402 (2004)

\bibitem{Koga-2005}
A. Koga, N. Kawakami, T. M. Rice, and M. Sigrist,
Phys. Rev. B {\bf 72}, 045128 (2005).
\bibitem{deMedici-2005}
L. de' Medici, A. Georges, and S. Biermann,
Phys. Rev. B {\bf 72}, 205124(2005).
\bibitem{Song-2005}
Y. Song and L.-J. Zou, Phys. Rev. B {\bf 72}, 085114 (2005).

\bibitem{deMedici-2009}
L. de' Medici, S. R. Hassan, M. Capone, and X. Dai,
Phys. Rev. Lett. {\bf 102}, 126401 (2009).
\bibitem{Song-2009}
Y. Song and L.-J. Zou,
Eur. Phys. J. B {\bf 72}, 59 (2009).
\bibitem{Jakobi-2013}
E. Jakobi, N. Bl\"{u}mer, and P. van Dongen,
Phys. Rv. B {\bf 87}, 205135 (2013).

\bibitem{YKNiu-2018}
Y. K. Niu, J. Sun, Y. Ni, and Y. Song,
Physica B {\bf 539}, 106 (2018).

\bibitem{Liebsch-2005}
A. Liebsch, Phys. Rev. Lett.
{\bf 95}, 116402 (2005).
\bibitem{deMedici-2011}
L. de' Medici, Phys. Rev. B {\bf 83}, 205112 (2011).
\bibitem{JSun-2015}
J. Sun, Y. Liu, and Y. Song,
Acta Phys. Sin. {\bf 64}, 247101 (2015).

\bibitem{Costi-2007}
T. A. Costi and A. Liebsch,
Phys. Rev. Lett. {\bf 99}, 236404 (2007).
\bibitem{Medici-2011}
L. de' Medici, J. Mravlje, and A. Georges,
Phys. Rev. Lett.{\bf 107}, 256401 (2011).
%

\bibitem{Greger-2013}
M. Greger, M. Kollar, and D. Vollhardt
Phys. Rev. Lett. {\bf 110}, 046403 (2013).

\bibitem{Fernandez-2018}
Y. N\'{u}\~{n}ez-Fern\'{a}ndez, G. Kotliar, and K. Hallberg,
Phys. Rev. B {\bf 97}, 121113(R) (2018).

\bibitem{XJHan-2016}
X.-J. Han, Y. Liu, Z.-Y. Liu, X. Li, J. Chen, H.-J. Liao,
Z.-Y. Xie, B. Normand, and T. Xiang,
New J. Phys. {\bf 18}, 103004 (2016).

\bibitem{Phillips-2010}
P. Phillips,
Rev. Mod. Phys. {\bf 82}, 1719 (2010).

\bibitem{Leigh-2009}
R. G. Leigh and P. Phillips,
Phys. Rev. B {\bf 79}, 245120 (2009).
%
\bibitem{Yamaji-2011}
Y. Yamaji and M. Imada,
Phys. Rev. B {\bf 83}, 214522 (2011).
%
\bibitem{Zhou-2014}
S. Zhou, Y. Wang, and Z. Wang,
Phys. Rev. B {\bf 89}, 195119 (2014).
%
\bibitem{Lee-2017}
S.-S. B. Lee, J. von Delft, and A. Weichselbaum,
Phys. Rev. Lett. {\bf 119}, 236402 (2017).
%
\bibitem{Lee-2-2017}
S.-S. B. Lee, J. von Delft, and A. Weichselbaum,
Phys. Rev. B {\bf 96}, 245106 (2017).


\bibitem{Nishimoto-2004}
S. Nishimoto, F. Gebhard, and E. Jeckelmann,
J. Phys. Condens. Matter {\bf 16}, 7063 (2004).
\bibitem{Gull-2010}
E. Gull, D. R. Reichman, and A. J. Millis,
Phys. Rev. B {\bf 82}, 075109 (2010).
\bibitem{Granath-2014}
M. Granath and J. Sch\"{o}tt,
Phys. Rev. B {\bf 90}, 235129 (2014).

\bibitem{Georges-1996}
A. Georges, G. Kotliar, W. Krauth, and M. J. Rozenberg,
Rev. Mod. Phys. {\bf 68}, 13 (1996).

\bibitem{Dagotto-1994}
E. Dagotto,
Rev. Mod. Phys. {\bf 66}, 763 (1994).


\bibitem{Caffarel-1994}
M. Caffarel and W. Krauth,
Phys. Rev. Lett. {\bf 72}, 1545 (1994).

\bibitem{Capone-2007}
M. Capone, L. de' Medici, and A. Georges,
Phys. Rev. B {\bf 76}, 245116 (2007).


\bibitem{Arita-2005}
R. Arita and K. Held,
Phys. Rev. B {\bf 72} 201102 (2005).







\end{thebibliography}

\end{document}